\pdfoutput=1
\documentclass[12pt]{iopart}

\usepackage{iopams}  
\expandafter\let\csname equation*\endcsname=\relax
\expandafter\let\csname endequation*\endcsname=\relax

\usepackage{physics}                            
\usepackage{bm}									
\usepackage{enumerate}							
\usepackage{amssymb}							
\usepackage[dvipsnames]{xcolor}					
\usepackage[colorlinks=true,allcolors=RoyalBlue]{hyperref}
\hypersetup{colorlinks=true}
\usepackage[capitalize]{cleveref}               
\usepackage{graphicx,color}
\usepackage{xcolor}
\usepackage{bbold}                              
\usepackage{cite}
\usepackage{tikz}
\usepackage[utf8]{inputenc}
\usepackage{braket}
\usepackage[normalem]{ulem} 
\usepackage{comment}
\usepackage{relsize}
\usepackage{cancel}
\usepackage{amsmath}
\usepackage{amssymb}
\usepackage{array}

\makeatletter
\renewcommand{\thefootnote}{\arabic{footnote}}
\def\@fnsymbol#1{\arabic{#1}}
\renewcommand\@makefnmark{\hbox{\textsuperscript{\thefootnote}}}
\long\def\@makefntext#1{\noindent\hbox{\textsuperscript{\thefootnote}}#1} 
\makeatother
\usepackage{etoolbox}
\makeatletter
\newrobustcmd{\fixappendix}{%
  \patchcmd{\l@section}{1.5em}{7em}{}{}%
  \patchcmd{\l@subsection}{2.3em}{7em}{}{}%
}
\makeatother

\begin{document}

\title[Covariant fermionic path integral]{A covariant fermionic path integral for scalar Langevin processes with multiplicative white noise}

\author{Daniel G.~Barci}
\address{Departamento de F\'{\i}sica Te\'orica, Universidade do Estado do Rio de Janeiro,
Rua S\~ao Francisco Xavier 524, 20550-013, Rio de Janeiro, RJ, Brazil.}

\author{Leticia F.~Cugliandolo}
\address{Laboratoire de Physique Théorique et Hautes Energies, CNRS-UMR 7589, Sorbonne Université, 4 Place Jussieu, 75252 Paris Cedex 05, France}

\author{Zochil Gonz\'alez Arenas}
\address{Departamento de Matem\'atica Aplicada, IME, Universidade do Estado do Rio de Janeiro, Rua S\~ao Francisco Xavier 524, 20550-013, Rio de Janeiro, RJ, Brazil}

\begin{abstract}
We revisit the construction of the fermionic path-integral representation of overdamped scalar Langevin processes with multiplicative white noise, focusing on the covariance of the generating functional under non-linear changes of variables. We identify the transformations of the auxiliary (commuting and anticommuting) variables that ensure covariance under such transformations. The subtleties induced by the non-differentiable trajectories of the stochastic dynamics are encoded in the fermionic statistics and in the regularization of its contact terms. 
Upon integrating out the auxiliary variables using the Stratonovich prescription, we recover the Onsager--Machlup action independently derived through a higher-order discretization scheme. This agreement provides a non-trivial consistency check of the continuum fermionic formulation, which is constructed without explicitly discretizing the Langevin equation.
The evaluation of the Grassmann integral requires a prescription for coincident-time
contributions. We show that the prescription used here, which drops these contributions, is equivalent to reading
the resulting Onsager--Machlup action with the higher-order discretization of that scheme and an end-point
normalization of the measure.
\end{abstract} 
\vspace{10pt}
\begin{indented}
\item[]\today
\end{indented}

\maketitle

\tableofcontents

\markboth{Covariant fermionic path integral}{Covariant fermionic path integral}

\setcounter{footnote}{0} 

\section{Introduction}
\label{sec:introduction}

White noise-driven Langevin processes provide a fundamental stochastic framework for modeling dynamical systems subject to random fluctuations~\cite{gardiner_handbook_1994,kampen_stochastic_2007,oksendal_stochastic_2013}. Originally rooted in statistical physics to describe Brownian motion, 
these processes have become a cornerstone across disciplines, enabling the analysis of, e.g., thermal fluctuations in chemical reactions via Kramers’ escape rate theory, 
the modeling of financial asset prices in quantitative finance, and the simulation of noisy biological signals in neural dynamics and gene expression. Their mathematical tractability, particularly through the associated Fokker-Planck equation for the probability distribution, allows for powerful predictions of stationary states, 
transition rates, and first-passage times.

Multiplicative noise models are exceptionally relevant for describing systems in strongly dissipative environments where the noise intensity varies with 
the variable of interest itself, such as diffusion in inhomogeneous media, population dynamics in stochastic environments with density-dependent noise, 
and the directional sensing of cells where signal amplification depends on the local concentration of chemoattractants. 
The state-dependent noise in these models can qualitatively alter the system's behavior, inducing noise-driven transitions, stabilizing 
otherwise unstable states, and producing phenomena like noise-induced ordering, which are absent in their additive noise counterparts~\cite{Lindenberg83,GarciaOjalvo99, Caravagna13,Volpe16,Sarkka19}.

The overdamped limit is an ubiquitous approximation in many physical and biological contexts. 
It is formally derived when inertial effects (modeled by a second-order derivative term) are negligible compared to frictional damping, 
leading to a first-order Langevin equation where the velocity is instantaneously slaved to the force and noise. 

Overdamped Langevin equations with white noise produce non-differentiable solutions. 
The stochastic calculus needed to treat such processes 
was developed long ago, notably by It\=o\cite{Ito1951} and Stratonovich\cite{Strato-original_1962,Stratonovich92}. In theoretical physics, 
it is often more convenient to use a generating functional formalism 
in which all possible trajectories are summed over with their corresponding 
probability weight. The generating functional takes, thus, the form of a path 
integral (see Refs.~\cite{zinn-justin_quantum_2002,chaichian_stochastic_2001,kleinert_path_2009,Thibaut-2022} for reviews)
over non-differentiable trajectories.  In the statistical physics contexts, these go under the names of 
Onsager--Machlup~\cite{Onsager1953,Onsager1953b} and Martin-Siggia-Rose-Jensen-deDominicis~\cite{MSR1973,Janssen-1976,deDominicis} representations. 
Of special utility, in particular to prove symmetries, 
is the fermionic formulation in which auxiliary Grassmann variables~\cite{Berezin1966} are introduced to 
represent a determinant associated to the change of variables from the noise to the 
stochastic variable itself in the functional integration.

Subtleties related to the action of non-linear changes of variables, 
linked to the non-differentiability of the trajectories, were noted in the very simple 
example of a Brownian particle moving in a two dimensional space 
by  Edwards and Gulyaev~\cite{edwards_path_1964}. 
This problem can obviously 
be treated in Cartesian or polar coordinates and, 
while one could safely change from one to the other at the level of the Langevin equations, 
a direct transformation within the path-integral formulation requires great care, due to subtleties 
associated with the non-differentiable Wiener paths.
Indeed, several authors from the statistical physics~\cite{graham_covariant_1977,graham_path_1977,deininghaus_nonlinear_1979,aron_dynamical_2016,Cugliandolo-Lecomte17a,Cugliandolo_2019,Thibaut-2022} 
and quantum field theory~\cite{dewitt_dynamical_1957,gervais_point_1976,Sa77,Langouche79,Langouche81,Tirapegui82,AlDa90,ApOr96,Ding2022} branches 
of theoretical physics were concerned with the definition of a path-integral which respects
covariance, that is, that maintains the same form when written in terms of 
different variables related to each other by non-linear changes of variables. 

Path integrals are more sensitive to discretization issues than Langevin equations --and this in spite of the fact that the white noise may have been integrated over and not appear explicitly in
the path-integral action. In~\cite{Cugliandolo_2019,Thibaut-2022}, 
a second order (in the stochastic variable increment 
during an infinitesimal time step) discretization scheme was used  to build such a covariant 
generating functional. In these references, the linear part of the increment was chosen  
to be of Stratonovich kind and this allowed one 
to  deal with functions in the path integrals, once the continuous time limit was taken, 
as if these were differentiable.
The extension to a generic prefactor of the linear term was studied in~\cite{Lecomte-unpublished}.
A complementary path-integral approach accounting explicitly
for different stochastic prescriptions and more general
noise statistics has recently been developed in
Ref.~\cite{AbrilBermudez2025}.

The natural question that emerges is whether the covariant generating functional can be obtained using the 
fermionic representation in a purely continuous-time approach~\cite{arenas2010,Miguel2015}.
In this paper, we reexamine the fermionic formulation of the path integral in a one dimensional case, 
ensuring that covariance properties are preserved at each step of the construction.
Once all auxiliary variables are integrated away, we favorably compare the resulting Onsager--Machlup expression with the one 
obtained using a higher-order discretization scheme and no fermions in~\cite{Cugliandolo_2019,Thibaut-2022}.

The paper is organized as follows. In Sec.~\ref{sec:Langevin} we present the Langevin equation 
with multiplicative white noise and we set the notation. In Sec.~\ref{subsec:construction} we recall the 
construction of the generating functional and, in particular, we use the Grassmann representation 
of the determinant. The basics of Grassmann calculus is recalled in~\ref{sec:Determinants}. 
Section~\ref{sec:covariance} provides the proof of covariance 
of the fermionic generating functional under non-linear changes of variables. 
In Sec.~\ref{integration} we integrate out the auxiliary and fermionic variables and 
we derive, in two alternative ways, the Onsager--Machlup generating functional. 
Finally, we present our conclusions and some lines for future research.

\section{Overdamped Langevin equations with multiplicative white noise}
\label{sec:Langevin}

In this Section we define an overdamped Langevin process subject to multiplicative white noise and we recall a number of its properties.
The presentation is useful to set the notation and conventions that we use in the rest of the paper.

\subsection{Definitions}

We consider the Langevin equation
\begin{equation}
 \frac{{\rm d}x(t)}{{\rm d}t} \stackrel{\alpha}{=}  f(x(t)) + g({x}(t)) \, \eta(t)  
 \; , 
  \label{eq:LangevSystem}
\end{equation}
 where   $ x $ is a function of time $t$ which takes real values, and  $\eta(t)$ is a  Gaussian white noise with mean and correlations
 given by 
\begin{equation}
 \left\langle \eta(t)\right\rangle   = 0 \; \mbox{,} 
 \qquad\;\;\;
\left\langle  \eta(t)  \eta(t')\right\rangle = \delta(t-t')
\; , 
\label{eq:whitenoise}
\end{equation}
for all $t$ and $t'$. 
The drift force, $f( x)$,  and  the diffusion coefficient, $g(x)$, 
are,  in principle,  arbitrary smooth functions of $x$. 
The noise intensity is included in the definition of the diffusion factor $g$. By replacing $g \to \sqrt{2 D}g$,  
we recover the notation of Refs.~\cite{Cugliandolo_2019,Thibaut-2022}. Alternatively, 
by replacing $g \to \sigma g $, we recover the notation of Refs.~\cite{MorenoBarciZochil-2019,MorenoBarciZochil-2020}.
The exact meaning of “$\alpha$” above the symbol “$=$” will be explained by the end of this section.

Because $\eta(t)$ is delta-correlated, the product $g(x(t)) \, \eta(t)$ in Eq.~(\ref{eq:LangevSystem}) is mathe\-ma\-ti\-cally ill-defined and requires a proper interpretation.
One way to face this problem is to consider the integral
\begin{equation}
\int    g( x(t))\, \eta(t) \, {\rm d}t= \int  g(x(t))\;  {\rm d}W(t) 
\; ,
\end{equation} 
where we have formally introduced the Wiener processes $W(t)=\int_{t_0}^t {\rm d}t' \, \eta(t')$. 
In a similar way to the Riemann-Stieltjes integral, this integral represents
\begin{eqnarray}
\int   g(x(t))\;  {\rm d}W(t) 
= \lim _{n\to\infty} \sum_{\ell=1}^n  g(x(\tau_\ell))(W(t_{\ell})-W(t_\ell-1))
\; ,
\label{eq:Wiener}
\end{eqnarray}
where the time integration domain has been discretized and $\tau_\ell$ is taken in the interval  $[t_{\ell-1},t_\ell]$ with $\ell=1,\ldots,n$.
The above  limit  is taken in the sense of a {\em mean-square limit}~\cite{gardiner_handbook_1994}.  
While a smooth $W(t)$ would yield a unique limit independently of $\tau_\ell$, $W(t)$ is in fact nowhere differentiable. 
White noise exhibits infinite variance and fluctuates infinitely often in any interval. As a result, the integral depends on the choice of  $\tau_\ell \in [t_{\ell-1},t_\ell]$, which is the origin of different stochastic prescriptions.

A comprehensive linear discretization approach is provided by the ``generalized Stratonovich prescription''~\cite{Hanggi1978} 
or ``$\alpha$-prescription''~\cite{Janssen-RG}, for which
\begin{eqnarray}
g(x(\tau_\ell)) &=& g(\alpha x(t_\ell) + (1-\alpha) x(t_{\ell-1}))  
\label{eq:prescription}
\end{eqnarray}
 with $0\le \alpha \le 1$.
 This formulation includes both the It\=o ($\alpha=0$), and Stratonovich ($\alpha=1/2$) schemes as particular instances. 
The post-point prescription, $\alpha=1$, is  known as the kinetic or the H\"anggi-Klimontovich interpretation.
The interpretation of Eq.~\eqref{eq:LangevSystem} follows the generalized Stratonovich prescription, indicated by the notation $\stackrel{\alpha}{=}$, thereby fixing the applicable stochastic calculus rules.

\subsection{Rules of calculus}
\label{sec:rules-of-calculus}

A key feature of stochastic integration is that each $\alpha$ value corresponds to a different calculus convention. Indeed, only the Stratonovich prescription ($\alpha=1/2$) preserves the ordinary rules of calculus.
This statement is made explicit through the chain rule derived for the $\alpha$-prescription.
Consider, for instance, an arbitrary function $H(x)$ of the stochastic variable $x$, 
satisfying Eq.~(\ref{eq:LangevSystem}) in the $\alpha$-prescription. Just by using a Taylor expansion one shows
 that
\begin{eqnarray}
\frac{{\rm d}H }{{\rm d}t}
&=&
\frac{\partial H}{\partial x} \, \frac{{\rm d}x}{{\rm d}t}+\left(\frac{1-2\alpha}{2}\right) g^2(x)
\,
\frac{\partial^2 H}{\partial x^2 }
\label{eq:Chain-Rule}
\end{eqnarray}
(see Ref.~\cite{Miguel2015} for a rigorous proof).
For the choice $\alpha=1/2$, Eq.~(\ref{eq:Chain-Rule}) is the usual chain rule. However, 
for $\alpha\neq 1/2$, an extra term proportional to the factor $g^2(x)$ is added.
This affects all other rules of calculus as, for instance, integration by parts. 

In Eq.~(\ref{eq:prescription}) the function $g(x)$ was evaluated at a linear combination of the past and future values 
which can also be written as
\begin{equation}
x(\tau_\ell) = x(t_{\ell-1}) + \alpha \Delta x  
\end{equation}
with
\begin{equation}
  \Delta x \equiv x(t_\ell) - x(t_{\ell-1}) 
\; . 
\end{equation}

The expansion in powers of $\Delta x $ can actually  be extended one more order
to use a second-order discretization scheme
\begin{equation}
x (\tau_\ell) 
=
x(t_{\ell-1}) + \alpha \Delta x + \beta_g(x)( \Delta x)^2 
\label{eq:covariant-M-discretization}
\end{equation}
or to even higher-orders. The subindex $g$ in the factor $\beta_g(x)$ stresses the fact that it can depend on the state-dependent diffusion
factor $g(x)$. This strategy has been commonly followed in the literature to increase the precision of 
numerical integration algorithms~\cite{Mannella02,Kloeden12}. In the vanishing time-step 
limit, $t_\ell - t_{\ell-1} \to 0$, the chain rule (\ref{eq:Chain-Rule}) is recovered with no extra contributions
from the higher order terms. However, when building the path integral generating functional with this 
discretization procedure, the second
order terms in Eq.~(\ref{eq:covariant-M-discretization}) yield a non-vanishing contribution even in the 
continuous time limit. As  shown in~\cite{Cugliandolo_2019,Thibaut-2022}, this contribution ensures  covariance
when the factor $\beta_g(x)$ is chosen to be 
\begin{equation}
\beta_g(x) = - \frac{1}{12} \frac{g'(x)}{g(x)} + \frac{1}{24} \frac{g''(x)}{g'(x)} 
\; , 
\label{eq:betag}
\end{equation}
where $g'$ means differentiation of $g$ with respect to $x$.

\subsection{Covariance}
\label{sec:covariance-1d-Langevin}

We adopt a broad notion of covariance: namely, that physical laws should be independent of the particular coordinates used to describe a system. In other words, the predictions of a theory should not depend on an arbitrary parametrization of the state space or on the observer’s frame of reference.
In the present work, this principle is applied to Langevin processes. Covariance guarantees that the stochastic dynamics are independent of the specific choice of variables or coordinates used to represent the underlying physical process.  A covariant Langevin equation preserves this principle by ensuring that the stochastic dynamics are represented consistently under coordinate transformations.

The transformation of multiplicative white noise Langevin equations under a generic invertible non-linear change of variables
is a non-trivial problem  initially studied by It\=o~\cite{Ito1951}. 
Let us illustrate this issue by  performing the following change of variables 
\begin{equation}
u(t)= U(x(t))
\label{eq:U}
\end{equation}
where $x$ satisfies the Langevin equation~(\ref{eq:LangevSystem}) and 
$U(x)$ is a smooth invertible function of $x$ in the sense that $x=U^{-1}(u)$ exists and it is also smooth.  
Then, by using the generalized chain-rule~(\ref{eq:Chain-Rule}) 
\begin{equation}
\frac{{\rm d}u}{{\rm d}t} =  U'(x)  \, \frac{{\rm d}x}{{\rm d}t} + \frac{(1-2\alpha)}{2}  \, g^2(x) \, U''(x)
\; , 
\end{equation}
and it is not difficult to show that $u(t)$ satisfies
\begin{eqnarray}
\frac{{\rm d}u}{{\rm d}t} & \stackrel{\alpha}{=} & F(u) + \, G(u) \, \eta(t)
\; , 
\label{eq:Langevinu-1v}
\end{eqnarray}
where 
\begin{eqnarray}
F(u)&=&  U'[U^{-1}(u)] \, f[U^{-1}(u)] \, + \, \frac{(1 - 2\alpha)}{2} \, g^2[U^{-1}(u)] \, U''[U^{-1}(u)] \; ,
\label{eq:F-main} 
\\
[4pt]
G(u) &=& U'[U^{-1}(u)] \, g[U^{-1}(u)]\; ,
\label{eq:G-main}
\end{eqnarray}
$U'(x)={\rm d}U/{\rm d}x$ and $U''(x)={\rm d}^2U/{\rm d}x^2$.  
Thus, the new stochastic variable $u(t)$ satisfies a Langevin equation with the same structure as the one for $x(t)$, 
with  the drift $F(u)$ and diffusion function $G(u)$, given by Eqs.~(\ref{eq:F-main}) and (\ref{eq:G-main}), respectively.
Using a linear or higher-order discretization scheme is irrelevant at  the level of the continuous time Langevin equation 
and its transformation. Only the linear parameter $\alpha$ (and not the function $\beta_g$) appears explicitly.
  
\section{Fermionic path integral formalism}
\label{sec:GeneratingFunctional}

In this section, we shift from the Langevin framework to the path integral formalism. 
The goal is to derive a covariant generating functional avoiding the explicit use of a discretization scheme, 
by adapting the techniques presented in Refs.~\cite{arenas2010,arenas2012,Arenas2012-2,Miguel2015}. 
We also update the notation in a way that will allow us to identify the covariant properties developed in the next section. 

\subsection{Construction}
\label{subsec:construction}

Here, closely following Ref.~\cite{arenas2010}, we derive the generating functional of correlation functions as a fermionic path integral.

We are interested in computing $n$-point correlation functions 
\begin{equation}
\left\langle x(t_1) \dots x(t_n) \right\rangle
\label{eq:correlation-function}
 \end{equation}
 where the average $\langle \dots\rangle$ is over white noise realizations (and possibly also initial conditions). To calculate these correlation functions we should know the $n^{\rm th}$-order joint probability function of the random variable $x$. 
A way to proceed is to solve the Langevin Eq.~(\ref{eq:LangevSystem}) and express the correlations in terms of the 
solution ${\bar x}_{[\eta]}(t)$ for a particular realization of the noise and certain initial condition $x(t_0)=x_0$:
\begin{equation}
\left\langle x(t_1) \dots x(t_n) \right\rangle 
\equiv 
 \left\langle \bar x_{[\eta]}(t_1) \dots \bar x_{[\eta]}(t_n) \right\rangle_{\eta}
 \; .
\end{equation}
The expression $\langle \, \ldots \, \rangle_{\eta} $ represents the average over the stochastic Gaussian noise $\eta$:
\begin{equation}
\langle \, \ldots \, \rangle_{\eta} \; = \int {\cal D}\eta \;  \ldots \; e^{-\frac{1}{2} \int {\rm d}t\ \eta^2 }
\; .
\end{equation}
The functional measures ${\cal D}\eta$  is a  short-hand notation for $\prod_{k=1}^K {\rm d}\eta(t_k)/\sqrt{2\pi}$ where the index $k$ labels the 
 times, $t_k = k \, \Delta t$,  at which the concerned time interval, $t_f-t_0$, is discretized.
The correlation functions (\ref{eq:correlation-function}) can then be obtained from the  generating functional
\begin{equation}
 Z[J] =  \left\langle e^{\; \mathop{\mathlarger{\int}} {\rm d}t\ J(t)  {\bar{x}}_{[\eta]}(t)}\right\rangle_{\eta}
 \; , 
\label{ZJxbar}
\end{equation}
by simply differentiating with respect to the 
source $J$,
\begin{displaymath}
 \left\langle \bar x_{[\eta]}(t_1) \dots \bar x_{[\eta]}(t_n) \right\rangle_{\eta}  = \frac{\delta ^n Z[J]}{\delta J(t_n) \dots \delta J(t_1) } \Bigg \vert_{J=0} 
 \!\!\!\!\! ,
 \!\!\!\!\!\! \ 
\end{displaymath}
and evaluating at $J(t)=0$.
Henceforth, all time integrals run from an initial time $t_0$ to a generic final time $t_f$ that we do not write 
explicitly.

Our goal is to find a functional representation of 
$Z[J]$ that avoids the need to explicitly solve the Langevin equation and does not require making
 the discretization scheme explicit.
For this purpose, we begin by introducing  a functional integral over $x(t)$ and a delta-functional which constraints 
its time-dependence to be a solution of the stochastic differential equation. 
Thus, we rewrite Eq.~(\ref{ZJxbar}) in the following form:
\begin{equation}
 Z[J] = {\left\langle \int {\cal D}x\; \delta[x(t)- { \bar{x}}_{[{\bf\eta}]}(t)] \ e^{\; \mathop{\mathlarger{\int}} {\rm d}t \; J(t)
  x(t) }\right\rangle }_{\eta}
\end{equation}
or, since the only noisy terms are contained in  ${ \bar x}_{[\eta]}(t)$,
\begin{equation}
 Z[J] =  \int {\cal D}x\;  e^{\; \mathop{\mathlarger{\int}} {\rm d}t \; J(t)
  x(t) }\; \left\langle\delta[x(t)- {\bar{x}}_{[{\bf\eta}]}(t)]\right\rangle_{\eta} \ .
 \label{eq:Zdeltaxeta}
\end{equation}
Here, the measure ${\cal D}x$ represents an integration over all paths starting at 
$x(0)=x_0$.
The Dirac delta functional imposes the validity of the Langevin equation at all times.
The advantage of this expression is that the source $J$ is no longer coupled to the solution of 
the stochastic differential equation.
 The next step is to eliminate the explicit dependence on this solution by using the following property of the delta-functional,
\begin{equation}
\delta[x(t)- { \bar{x}}_{[{\eta}]}(t)]=  \delta[{\hat{O}_{[\eta]}}(x(t))]\;{\rm det}\left(\frac{\delta \hat{O}_{[\eta]}(x(t))}{\delta x(t')}\right)
\; ,  
\label{delta-property}
\end{equation}
 where ${ \bar{x}}_{[\eta]}(t)$ is a root of ${\hat{O}_{[\eta]}}(x)$, \emph{i.e.}, 
${ \hat{O}_{[\eta]}}({ \bar{x}}_{[\eta]}(t)) = 0$.
We assume here that the operator $\hat {O}$ has only one root. In other words, we are supposing that, 
for a particular realization of the noise, the trajectory is completely specified by the initial condition $x(t_0)=x_0$.
The form of the operator ${\hat{O}_{[\eta]}}(x)$ is not uniquely determined since there are different ways in which one 
can write the Langevin equation. A simple  choice is 
\begin{equation}
 \hat{O}_{[\eta]}(x(t)) = \frac{{\rm d}x(t)}{{\rm d}t} - f(x) - g(x)\eta (t)
 \; .
 \label{eq:O}
\end{equation} 
Correspondingly, the differential operator $\delta \hat{O}_{[\eta]}(x(t))/\delta x(t')$ is
\begin{eqnarray}
\frac{\delta \hat{O}_{[\eta]}(x(t))}{\delta x(t')} 
&=& 
\left[  \frac{{\rm d}}{{\rm d}t} - f'(x) - g'(x)\eta (t)\right]   \times \, \delta (t-t')
\; ,
\label{eq:O'}
\end{eqnarray}
where $f'$ indicates differentiation of $f$ with respect to $x$.
Substituting Eq.~(\ref{delta-property})  into Eq.~(\ref{eq:Zdeltaxeta}), the generating functional now reads, 
\begin{equation}
 Z[J] =  \int {\cal D}x \; e^{\; \mathop{\mathlarger{\int}}  {\rm d}t \; J(t)
  x(t) }\; \left\langle\delta[{ \hat{O}_{[\eta]}}(t)] \det\left(\frac{\delta \hat{O}_{[\eta]}(t)}{\delta x(t')}  \right)\right\rangle_{\eta}  \; , 
\label{eq:Z2}
\end{equation}
with the definitions of ${ \hat O}_{[\eta]}$ and $\delta \hat O_{[\eta]}/\delta x$ given by Eqs.~(\ref{eq:O}) and~(\ref{eq:O'}), 
respectively. At this point, we have completely eliminated from the generating functional any reference to the explicit solution of the stochastic differential equation. 

The main difficulty now is   to correctly obtain the statistical mean value over the noise $\eta$ appearing in Eq.~(\ref{eq:Z2}).  In the multiplicative noise 
case, not only the delta-functional but also the determinant are $\eta$--dependent. As a consequence, for the aforementioned purpose, we will 
represent the delta-functional and the determinant through integral expressions using auxiliary variables.  
Introducing an auxiliary time-dependent function
 ${\varphi}(t)$ we can represent the delta-functional as a Fourier functional integral in the following form, 
\begin{equation}
\delta[ {\hat O}_{[\eta]}(x) ] = \int {\cal D} {\varphi} \;  
 e^{- {\rm i} \, \mathop{\mathlarger{\int}}  {\rm d}t \  \varphi(t) \, \left[ \dot x(t) - f(x(t)) - g(x(t))\eta(t) \right] }
\label{delta}
\end{equation}
where the ``dot'' means time differentiation. 

The representation of the determinant is more involved. In the same way that inverse determinants can be represented by Gaussian integrals, a determinant itself can be represented as a Gaussian integral over anti-commuting variables (see \ref{sec:Determinants}). 
Thus, we introduce a couple of  functions of time 
$ {\xi}(t) $ and  $\bar{{\xi}}(t) $, which obey the Grassmann algebra
\begin{equation}
\{\xi(t),\xi(t')\}=\{\bar\xi(t),\bar\xi(t')\}=\{\xi(t),\bar\xi(t')\}=0
\; ,
\end{equation}
where $\{ \, , \}$ represents an anti-commutator. 
These relations imply, in particular, $(\xi(t))^2=(\bar\xi(t))^2=0$. In terms of these auxiliary variables, 
the determinant can be written as 
\begin{eqnarray}
&& 
\det \left(\frac{\delta \hat{O}_{[\eta]}(x(t))}{\delta x(t')} \right) 
=
\int {\cal D} {\xi} {\cal D} \bar{{\xi}}
   \ e^{ \, \mathop{\mathlarger{\int}}  {\rm d}t  {\rm d}t'\;\bar{\xi}(t) \left(\frac{\delta \hat{O}_{[\eta]}(x(t))}{\delta x(t')} \right)\xi(t') }
   \nonumber  \\
 && 
 \qquad = 
 \int  {\cal D} {\xi} {\cal D} \bar{{\xi}}  
 \ e^{ \, \mathop{\mathlarger{\int}}  {\rm d}t \; \bar{\xi}(t) \dot\xi(t) \; -  \,  \mathop{\mathlarger{\int}}    {\rm d}t \; \bar{\xi}(t) \, \left[f'(x(t)) - g'(x(t)) \eta(t) \right] \, \xi(t)} 
\label{detgrassman}
\end{eqnarray}
(for mathematical details see Ref.~\cite{zinn-justin_quantum_2002} and \ref{sec:Determinants}).

Substituting the representations~(\ref{delta}) and~(\ref{detgrassman})  into Eq.~(\ref{delta-property}), we find
\begin{eqnarray}
&& \left\langle \delta [x(t)- { \bar{x}}_{{[\eta]}}(t)]\right\rangle _{\eta} 
\nonumber\\
&&
\qquad
\! = \!
\int {\cal D} {\xi} {\cal D} \bar{{\xi}} {\cal D }{\varphi}
\, 
 e^{ \, \mathop{\mathlarger{\int}}   {\rm d}t\;
\left[  \bar{\xi} \dot\xi -\bar{\xi} 
f' \xi  - {\rm i}  \varphi \left(\dot x - f \right) \right]
}
 \left\langle  e^{- \, \mathop{\mathlarger{\int}}  {\rm d}t\;
\left( \bar{\xi} g' \xi  - {\rm i} \varphi g \right) \eta}\right\rangle _{\eta} 
\; , 
\label{delta-exvalue}
\end{eqnarray}
where here and in the rest of this section we do not write the time and $x$ dependence in the 
exponentials to lighten the notation.
Since the noise distribution is Gaussian, it is now immediate to compute the average over the noise. We find,
\begin{eqnarray}
\left\langle  e^{-\, \mathop{\mathlarger{\int}}    {\rm d}t\;
\left( \bar{\xi} g'  \xi - {\rm i} \varphi g\right) \eta }\right\rangle _{\eta} 
 \propto  
e^{ \frac{1}{2} \, \mathop{\mathlarger{\int}}    {\rm d}t \;
\left[  -
2 {\rm i} \varphi g g' \bar{\xi} \xi 
- 
 (\varphi g)^2 \right] } 
\; .
\label{average}
\end{eqnarray}

Finally, introducing Eq.~(\ref{delta-exvalue}) into Eq.~(\ref{eq:Z2}) and using Eq.~(\ref{average}), we obtain the sought representation for the generating functional of the correlation functions,
\begin{equation}
 \mathit{Z}[J] = \int {\cal D}x {\cal D} {\varphi} {\cal D} {\xi} {\cal D}\bar{{\xi}} \;\; e^{-S[x, {\varphi}, \bar{{\xi}}, {\xi}] +   \mathop{\mathlarger{\int}} {\rm d}t \; J  x},
\label{eq:Z}
\end{equation}
where the ``action'' $S$ is given by
\begin{eqnarray}
S[x, {\varphi}, \bar{{\xi}}, {\xi}] =  
 \int   {\rm d}t 
 \left[ {\rm  i} \varphi \left( \dot{x} - f  + gg'  \bar{\xi}\xi \right)  
 + 
 \frac{1}{2} (g \varphi)^2   
  \right. 
- \left.\bar{\xi} \dot{\xi} +  f' \bar{\xi} \xi
 \right] .
 \;\;\;\;\;\;\;
\label{eq:action}
\end{eqnarray}
This expression summarizes the main goal of this section. 
It states that we can represent a stochastic multiplicative process as a functional integral over a set of commuting and anti-commuting variables. The multiplicative character of the process can be visualized 
through terms proportional to  $g'$. In the first term in Eq.~(\ref{eq:action}), 
the response variables ($\varphi$) couples to the Grassmann variables ($\bar\xi  \xi$)
through $gg'$.  This is a  typical signature of multiplicative noise  processes. 

Although the formal continuum fermionic action does not explicitly depend on the stochastic prescription parameter $\alpha$, the corresponding functional integral requires additional prescriptions in order to perform explicit calculations. In particular, the integration over the Grassmann variables requires an ultraviolet regularization, since the equal-time value of the retarded fermionic Green function, $G_R(0)$, is not determined by the formal continuum action. The relation between this regularization and the stochastic prescription for general values of $\alpha$ was discussed in Refs. [39,40]. However, a consistent treatment also requires specifying the rules of calculus to be used when manipulating the remaining commuting variables within the functional integral. The implementation of these rules for general stochastic prescriptions deserves further investigation.
Moreover, specifying $G_R(0)$ does not fully determine the
Grassmann integral. A separate prescription is needed for the
coincident-time contact terms that appear when the Grassmann series is re-exponentiated (see Sec.~\ref{subsubsec:first-route}).

In the present work, we restrict our analysis to the choice
$G_R(0)=1/2$, discard the contact terms ($\Theta_2=0$),
and use the ordinary rules of calculus.
With these assumptions, integrating out the auxiliary fields recovers
the covariant Onsager--Machlup action independently derived
using a higher-order discretization scheme adapted to non-linear
transformations. This agreement provides a non-trivial consistency
check of the continuum formulation. The explicit covariance proof
presented in the following section is also restricted to this
prescription.

\subsection{Covariance}
\label{sec:covariance}

The covariance of the Langevin process discussed in Sec.~\ref{sec:covariance-1d-Langevin} 
should be preserved  in the fermionic path integral formulation.
In this subsection we identify the transformations of the auxiliary and Grassmann variables used to express the 
determinant, which ensure the covariance of the  generating 
functional under a generic invertible non-linear transformation of the  stochastic variable. 
 We use the Stratonovich rule for the transformation of the variables and we assume that we  can use conventional calculus\footnote{The construction of the action and measure has been fully done in continuous time. 
 Still, we associate the term ``Stratonovich'' to the conventional differential calculus that we will use, 
 since this is the calculus that arises in the continuous time limit of the Stratonovich discretization rule.}
in the continuous time representation of the path integral as well.

The problem can be formulated in the following way. 
The correlation functions of a single stochastic variable $x$ can be derived from the generating functional
\begin{equation}
Z[J] =\int  {\cal D}x {\cal D}\varphi {\cal D}\bar\xi {\cal D}\xi \; e^{-S[x,\varphi,\bar\xi,\xi] + \, \mathop{\mathlarger{\int}} {\rm d}t \, J x } 
\label{eq:Zx}
\end{equation}
with the action given in Eq.~(\ref{eq:action}). 
Explicitly writing the temporal and variable depen\-den\-cies, the action reads
\begin{eqnarray}
&& S[x,\varphi,\bar\xi,\xi] =
  \int  {\rm d}t \; \Big{\{}  {\rm i} \varphi(t) \left[ \dot{x}(t) - f(x)  + g(x(t)) g'(x(t))\bar{\xi}(t)\xi(t) \right] \nonumber \\
 && 
 \quad \quad \qquad \quad  \qquad \quad 
 +  \frac{1}{2}g^2(x(t))\varphi^2(t) -\bar{\xi}(t) \dot{\xi}(t) + f'(x(t)) \bar{\xi}(t) \xi(t) \Big{\}} 
 \, .
\label{eq:actionx-1v}
\end{eqnarray}
On the other hand, if we make the change of variables $u=U(x)$ and 
we start from the Langevin equation~(\ref{eq:Langevinu-1v}), 
we should find a generating functional and an action with this very
 same structure but written in the new variables, 
\begin{equation}
\tilde Z[\tilde J]
=\int  {\cal D}u {\cal D}\tilde\varphi {\cal D}\bar\zeta {\cal D}\zeta \; e^{-\tilde S[u,\tilde\varphi,\bar\zeta,\zeta] +  \, \mathop{\mathlarger{\int}} {\rm d}t \, \tilde J u} 
\label{eq:Zu}
\end{equation}
with the action
\begin{eqnarray}
&& 
\tilde S[u,\tilde\varphi,\bar\zeta,\zeta]   =
  \int  {\rm d}t \; \Big{\{}  {\rm i} \tilde\varphi(t) \left[ \dot{u}(t) - F(u(t))  + G(u(t)) G'(u(t))\bar{\zeta}(t)\zeta(t) \right] 
  \nonumber\\
 && \quad   \qquad \quad  \qquad \quad 
 +\left. \frac{1}{2}G^2(u(t))\tilde\varphi^2(t) -\bar{\zeta}(t) \dot{\zeta}(t) + F'(u(t)) \bar{\zeta}(t) \zeta(t) \right\} ,
 \;\;\;\;\;\;\;\;
\label{eq:actionu-1v}
\end{eqnarray}
with the functionals $F(u(t))$ and $G(u(t))$ given by Eqs.~(\ref{eq:F-main}) and (\ref{eq:G-main}), respectively. In this case, the 
primes represent derivatives with respect to the argument $u$.

The question is whether one can transform $Z[J]$ in $\tilde Z[\tilde J]$, and vice versa, 
by performing the change of variables from $x$ to $u$, and the inverse. 
Since the auxiliary variable $\varphi$ and the fermionic ones $\xi$ and $\bar\xi$
 appear in $Z[J]$ we will have to define their transformation rules.
To check the covariance, we will start from Eq.~(\ref{eq:actionu-1v}) and we  
will rewrite $\tilde S[u,\tilde\varphi,\bar\zeta,\zeta]$   in terms of $x, \varphi, \bar\xi, \xi$.

First, we use Eqs.~(\ref{eq:F-main}) and~(\ref{eq:G-main}) which relate $F(u)$ and $G(u)$ to $f(x)$ and $g(x)$, 
to derive expressions for their derivatives with respect to $u$ again in terms of functions of~$x$:
\begin{eqnarray}
F'(u) &\equiv& 
\frac{dF(u)}{du} = \frac{d}{du} \{ U'[U^{-1}(u)] \, f[U^{-1}(u)] \}
\nonumber\\
&=& \frac{d}{dx} [U'(x) f(x)] \; \frac{dx}{du}
\nonumber\\
&=& [U''(x) f(x) + U'(x) f'(x)]  \; \frac{1}{U'(x)} 
\; , 
\label{eq:Fprime}
\\
G'(u) &\equiv& 
\frac{dG(u)}{du} =
 [U''(x) g(x) + U'(x) g'(x)]  \; \frac{1}{U'(x)} 
\; ,
\label{eq:Gprime}
\end{eqnarray}
where we did not write the time dependence explicitly.

Next, we propose the following  linear changes of variables for the 
set $\varphi, \bar\zeta, \zeta$,
\begin{align}
\tilde\varphi&=\frac{1}{U'(x)}\left( \varphi- {\rm i} \frac{U''(x)}{U'(x)} \, \bar\xi\xi\right)
\; , 
\label{eq:tildevarphi}
\\
\zeta&= U'(x) \, \xi 
\; , 
\label{eq:eta}
\\
\bar\zeta&= \frac{1}{U'(x)}\, \bar\xi
\; ,
\label{eq:bareta}
\end{align}
which conveniently satisfy $\bar\zeta \zeta =\bar \xi \xi$. 

Here, covariance refers to the invariance of the complete generating functional, including both the action and the functional measure, under the non-linear transformation of the stochastic variable and the accompanying transformations of the auxiliary fields given in Eqs. (39)–(41). The proof is carried out directly in continuous time, 
using the ordinary chain rule together with these field transformations. No explicit midpoint time discretization is introduced. Accordingly, we apply the standard chain rule,
\begin{align}
\dot u& = U'(x) \, \dot x 
\; , 
\\
\frac{{\rm d}[U'(x) \xi]}{{\rm d}t} &= U''(x) \,\dot x \xi + U'(x) \, \dot \xi
\; . 
\end{align}
With these transformations the integrand in the action becomes
\begin{eqnarray}
&&  {\rm i} \tilde\varphi(t) \left[ \dot{u}(t) - F(u(t))  + G(u(t)) G'(u(t)) \, \bar{\zeta}(t)\zeta(t) \right] + \frac{1}{2}G^2(u(t))\tilde\varphi^2(t) 
\nonumber \\
&&
\quad -\bar{\zeta}(t) \dot{\zeta}(t)  + F'(u(t)) \, \bar{\zeta}(t) \zeta(t)
\nonumber \\
&&
 = 
{\rm i} \varphi(t)
\left(\dot x(t) - f (x(t))+ g(x(t)) g'(x(t))  \, \bar \xi(t) \xi(t) \right) + \frac{1}{2} g^2(x(t))  \varphi^2(t) 
\nonumber\\
&& 
\quad - \bar \xi(t) \dot \xi(t) + f' (x(t)) \, \bar \xi (t) \xi (t)
+ \;  \frac{1}{2} [g'(x(t))]^2 \;
 (\bar \xi(t) \xi(t))^2
 \; .
\end{eqnarray}
The new term proportional to $(\bar \xi \xi)^2$ arises 
from the transformation of $\tilde\varphi$ and ${\tilde\varphi}^2$ but 
it vanishes identically thanks to the anti-commuting properties of the 
Grassmann variables.

Finally, we obtain 
\begin{equation}
\tilde S[u,\tilde\varphi,\bar\zeta,\zeta]= S[x,\varphi,\bar\xi,\xi] 
\; . 
\label{eq:covariant-action}
\end{equation}

We note that had we started with a process $u$ with additive noise, that is $G(u)=1$, after the transformation
$u=U(x)$, we would have ended with the action for a multiplicative noise process with 
$f=F/U'$ and $g=1/U'$. 

In order to prove the full covariance of the generating functional, we need to examine the 
transformation of the integration measure. The transformation of 
${\cal D} u{\cal D}\tilde\varphi{\cal D}\bar\zeta{\cal D}\zeta$ into 
${\cal D}x{\cal D}\varphi{\cal D}\bar\xi{\cal D}\xi$ involves a Jacobian, determined by the 
determinant of a $4\times 4$ (times-dependent) matrix. Exploiting the fact that 
$\partial u/\partial\varphi=\partial u/\partial \xi = \partial u / \partial \bar\xi =
\partial \bar \zeta / \partial \varphi = \partial \bar \zeta /\partial \xi = 
\partial \zeta / \partial \varphi =0$, and using 
$\partial u/\partial x = U'(x)$, $\partial \tilde \varphi/\partial \varphi = 1/U'(x)$, 
$\partial \zeta/\partial \xi = U'(x)$ and $\partial \bar \zeta/\partial \bar\xi = 1/U'(x)$, 
one shows that the 
Jacobian equals one for the change of integration variables proposed. The measure
\begin{equation} 
{\cal D} u{\cal D}\tilde\varphi{\cal D}\bar\zeta{\cal D}\zeta=
{\cal D} x {\cal D}\varphi{\cal D}\bar\xi{\cal D}\xi
\label{eq:covariant-measure}
\end{equation}
is indeed invariant. 

Therefore, 
$\tilde Z[\tilde J]=Z[J]$ after applying an adequate transformation of the source.
Thus,  the path integral formalism in the fermionic representation has the same covariance properties 
that the original Langevin equation.  

In Ref.~\cite{Ding2022}, a model-independent criterion for validating time-sliced path-integral representations of Langevin processes has been  proposed, based on the first three moments of the stochastic increment  $\Delta x$. Its application to a discretization of the present fermionic  action is an interesting question that we leave for future work.

In Sec.~\ref{integration} we will show how to recover,  after integrating out the fermions, the Onsager--Machlup 
action and measure which were deduced in Ref.~\cite{Cugliandolo_2019} for the $d=1$ problem.

\section{The Onsager--Machlup formulation}
\label{integration}
The convenience of the fermionic representation of the path integral lies in the fact that a wide range of symmetries,
notably supersymmetric transformations, can be implemented as linear transformations. 
However, for explicit calculations it is often more convenient to work with a representation written solely in terms of the original variable $x$.  
To obtain this form, the response field ${ \varphi}$ and the Grassmann variables $\{\bar{{\xi}}, {\xi}\}$ must be integrated out.  
This procedure involves several subtleties. In this section, we  provide a correct derivation of the one-dimensional Onsager--Machlup action. 
This analysis updates and refines the results previously reported in Refs.~\cite{arenas2010,arenas2012,Arenas2012-2}. 
We also show that our result coincides with the one communicated in  Ref.~\cite{Cugliandolo_2019}, which was 
obtained by means of a higher order discretization procedure.

The generating functional and the action are given by  Eq.~(\ref{eq:Zx}) and Eq.~(\ref{eq:actionx-1v}), respectively.
We proceed in two alternative ways, which yield the same result, though have to be followed with 
special care, as explained below.

\subsection{Order of integration: response variable followed by Grassmann variables}
\label{subsubsec:first-route}

As the $\varphi$ integral is Gaussian,
\begin{align}
I_\varphi\equiv\int  {\cal D}\varphi \;
e^{-\, \mathop{\mathlarger{\int}} {\rm d}t 
\; \left[ \frac{1}{2} g^2\varphi^2  +  {\rm i} \varphi \left( \dot{x} - f  + g g'\bar{\xi}\xi \right)\right]} \;,
\nonumber 
\end{align}
it can be performed exactly using standard methods. One obtains
\begin{align}
I_\varphi=({\det}(g^2))^{-1/2} \; e^{- \, \mathop{\mathlarger{\int}}  {\rm d}t 
\left[\frac{\left(\dot{x} - f\right)^2}{2g^2}  + \left(\dot{x}-f\right)\frac{g'}{g}\bar{\xi}\xi \right]} 
\; ,
\end{align}
where we have used the fact that $(\bar\xi(t)\xi(t))^2=0$ for all $t$.
The factor $(\det(g^2))^{-1/2}$ arises from normalizing the Gaussian integrals over $\varphi$ (ignoring numerical $2\pi$ factors).
 We retain this compact, implicit form throughout. Returning to the generating functional expression, we have
\begin{equation}
Z=\int {\cal D}x \left(\det(g^2)\right)^{-1/2} 
\; 
e^{-\, \mathop{\mathlarger{\int}}  {\rm d}t \; 
\frac{\left(\dot{x} - f\right)^2}{2g^2} } 
\; 
I_\xi(x)
\label{eq:ZIxi}
\end{equation}
with the definition
\begin{equation}
I_\xi(x)\equiv \int {\cal D}\bar\xi{\cal D}\xi \; e^{\, \mathop{\mathlarger{\int}} {\rm d}t
\; \left( \bar\xi\dot\xi-k\bar\xi\xi\right) }
\label{eq:Ixi}
\end{equation}
where $k$ is a function of $x(t)$,
\begin{equation}
k(x(t)) \equiv f'(x(t))+\left(\dot x(t)-f(x(t))\right)\frac{g'(x(t))}{g(x(t))}
\; .
\label{eq:kdef1}
\end{equation}
The next step  is to perform the integration over the Grassmann variables  in Eq.~(\ref{eq:Ixi}), 
using standard techniques of Grassmann path-integral calculus~\cite{zinn-justin_quantum_2002}.
We expand  the exponential of the second term, proportional to $k(x)$, in Taylor series:
\begin{align}
I_\xi(x)
= & \int {\cal D}\bar\xi{\cal D}\xi \; e^{\, \int {\rm d}t \; \bar\xi(t) \dot\xi(t)}  
\; \, \Big\{ 1-\, \mathop{\mathlarger{\int}} {\rm d}t \; k(x(t))\bar\xi(t)\xi(t) 
\nonumber \\
& \qquad + 
\frac{1}{2}\int {\rm d}t {\rm d}t'  \; k(x(t))k(x(t'))\bar\xi(t)\xi(t)\bar\xi(t')\xi(t')
+ \ldots \Big{\}}
\nonumber\\
= & \;\;
\langle 1 \rangle -\int {\rm d}t \; k(x(t))\left\langle\bar\xi(t)\xi(t)\right\rangle
\nonumber \\
&
\qquad
+\frac{1}{2}\int {\rm d}t {\rm d}t'  \; k(x(t))k(x(t'))\left\langle\bar\xi(t)\xi(t)\bar\xi(t')\xi(t')\right\rangle
+ \ldots
\end{align}
where the correlation functions $\langle\dots\rangle$ are computed using the Gaussian weight 
$\exp\int {\rm d}t \, \bar{\xi}(t)\dot{\xi}(t)$.

The average of $1$ in the first term is simply
\begin{equation}
 \int {\mathcal D}\bar\xi {\mathcal D}\xi \; e^{\int dt \, \bar\xi(t) \dot \xi(t)} = {\rm det} ( {\rm d}/{\rm d}t )
 \; . 
 \end{equation}
 
 The two-point correlation function is given by the Green's function of the first derivative operator ${\rm d}/{\rm d}t$ times 
 the normalization, that is, 
\begin{equation}
\left\langle \bar{\xi}(t) \xi(t')\right\rangle =  {\rm det} ( {\rm d}/{\rm d}t ) \; G(t,t') \; ,
\end{equation}
where $G(t,t')$ satisfies
\begin{equation}
\frac{{\rm d} G(t,t')}{{\rm d}t}=\delta(t-t')
\; .
\end{equation}
As usual, we need a prescription to compute the Green's function, generally given by the choice of 
initial conditions. For causality reasons we decided to work with the retarded Green's function, 
for which $G_R(t,t')=0$ if $t<t'$. In such a case, 
\begin{equation}
G_R(t,t')=\Theta(t-t')
\; ,
\label{eq:GR}
\end{equation}
where $\Theta(t)$ is the Heaviside distribution.
We will fix the value of $\Theta(0)$ below.

Next, we use Wick's theorem to factorize the multi-point correlation functions in terms of two-point ones, 
\begin{eqnarray}
\left\langle \bar{\xi}(t) \xi(t) \bar{\xi}(t') \xi(t')
\right\rangle & =&  \left\langle \bar{\xi}(t) \xi(t)
\right\rangle \left\langle \bar{\xi}(t') \xi(t') \right\rangle
- \left\langle \bar{\xi}(t)
\xi(t') \right\rangle \left\langle \bar\xi(t') {\xi}(t)
\right\rangle\, ,
\label{eq:WickGras}
\end{eqnarray}
where we have used the fact that $\langle\xi(t)\xi(t')\rangle=\langle\bar\xi(t)\bar\xi(t')\rangle=0$
for all $t$ and $t'$, and similarly for higher order terms.

With this, the integral $I_\xi$ takes the form, 
\begin{align}
I_\xi(x)= 
& 
\det({\rm d}/{\rm d}t) \, 
\left\{1-G_R(0) \int {\rm d}t \; k(x(t)) \right. 
\nonumber\\
&
\qquad\qquad\quad +\frac{1}{2!} \int {\rm d}t {\rm d}t'  \; k(x(t))k(x(t'))
\;  \left[ G_R^2(0)-G_R(t-t')G_R(t'-t) \right]
\nonumber \\
&
\qquad\qquad\quad  \left. +\frac{1}{3!}\int {\rm d}t {\rm d}t' {\rm d}t''\ldots\right\}
\; . 
\label{eq:Ixi2}
\end{align}
At this point, it is worth noting that the expression
between square brackets takes the values 
\begin{eqnarray}
&& 
\left[ G_R^2(0)-G_R(t-t')G_R(t'-t) \right] =
\nonumber\\
[5pt]
&& 
\qquad\qquad \qquad 
=
\left\{ 
\begin{array}{lcl}
G_R^2(0) & \qquad \mbox{for} & t\neq t' \\
& & \\
0  & \qquad \mbox{for} & t= t' 
\end{array}
\right.
\label{eq:GR2}
\end{eqnarray}
This is a consequence of the retarded character of the Green function. 
Also, this expression keeps track of the anti-symmetry of the Grassmann variables, since $\left\langle \bar{\xi}(t) \xi(t) \bar{\xi}(t) \xi(t)
\right\rangle=0$ (see Eq.~(\ref{eq:WickGras})). 

The coincident-time contributions, $t=t'$, require a separate discussion. Because the
bracket vanishes at $t=t'$, the second-order term is not $\frac12\big(G_R(0)\int{\rm d}t\,k(t)\big)^2$: it
differs from it by the coincident-time part of this square. This part is not negligible: $k(x(t))$
depends on $\dot x(t)$ which, on the trajectories that dominate the path integral, behaves like white
noise, so that $k(t)k(t')$ behaves like $g'^2(t)\,\delta(t-t')$. It produces the local contribution
\begin{equation}
\frac{\Theta_2}{2}\int{\rm d}t\,g'(x(t))^2
\label{eq:theta2}
\end{equation}
to the action, with $\Theta_2=G_R^2(0)$, i.e., $\Theta_2=1/4$ for $G_R(0)=1/2$. Keeping it leads to the
Stratonovich-discretized action, which is not covariant. In this paper we adopt instead the
regularization $\Theta_2=0$: we discard the coincident-time contribution of the $k$-dependent terms
and re-exponentiate the series in full. With this regularization the resulting Onsager--Machlup
action is covariant, while any $\Theta_2\neq0$ would add the term in Eq.~(\ref{eq:theta2}),
which is not invariant under changes of variables. The regularization concerns only the contact terms
of the $k$-dependent series. It does not modify the Grassmann correlation functions, which vanish at
coincident times, a property used in Sec.~\ref{subsubsec:second-route}. We then have,
 \begin{align}
I_\xi(x) = &
\det({\rm d}/{\rm d}t) \,
\left\{1-G_R(0)\int {\rm d}t \; k(x(t))\right. 
\left. + \, \frac{1}{2!} \left(G_R(0)\int {\rm d}t \; k(x(t))\right)^2\right.
\nonumber \\
&\qquad\qquad\quad  \left.-\frac{1}{3!}\left(G_R(0)\int {\rm d}t \; k(x(t))\right)^3+\frac{1}{4!}\ldots\right\}
\end{align}
and we can re-exponentiate the expression between curly brackets to deduce 
 \begin{align}
&I_\xi(x)=
\det({\rm d}/{\rm d}t) \,
\; e^{- \, G_R(0)  \, \mathop{\mathlarger{\int}} {\rm d}t \; k(x(t))}
\; . 
\end{align}
The Green function at the origin is not uniquely  defined.  Actually, the relation to the 
underlying discretization rule used to define the stochastic process, is encoded in the 
choice of $G_R(0)$ together with the regularization of the contact terms discussed above.
We fix  
\begin{equation}
G_R(0)=\alpha  = \frac{1}{2} \, ,
\end{equation} 
where $\alpha$ labels the linear term of the stochastic prescription~\cite{Janssen-RG,arenas2012, Arenas2012-2,Miguel2015}
which has been fixed to $1/2$.
Therefore, the integration over Grassmann variables takes the form, 
 \begin{align}
&I_\xi(x)=
\det({\rm d}/{\rm d}t) \, 
 \; e^{-\frac{1}{2} \, \mathop{\mathlarger{\int}} {\rm d}t \;  \left[ f'(x(t))+\left(\dot x(t)-f(x(t))\right)\frac{g'(x(t))}{g(x(t))}\right]}   \; .
\end{align}

Replacing the last expression for $I_\xi(x)$ in Eq.~(\ref{eq:ZIxi}) we finally find
\begin{equation}
Z=\int {\cal D}x \left(\det(g^2)\right)^{-1/2} e^{-S[x]}
\label{eq:ZOM}
\end{equation}
where the action is given by
\begin{align}
S[x]=&\int  {\rm d}t \; \left\{ \frac{1}{2}\left[ \frac{\dot x-f(x)}{g(x)}\right]^2 
+\frac{1}{2}\left[f'(x)  -f(x) \frac{g'(x)}{g(x)}\right] + \frac{1}{2} \dot x \, \frac{g'(x)}{g(x)} \right\}
\; . 
\label{eq:Sx-1v}
\end{align}
We have not written the $t$ dependence explicitly.
The last term in this equation can be recast in the form of the total derivative of
$(1/2)\ln(g^2)$ and integrated away, giving a boundary term.
We then find the desired result 
\begin{equation}
S[x] 
=
\frac{1}{2} \int  {\rm d}t \;  
\left\{
\left[ \frac{\dot x-f(x)}{g(x)}\right]^2 
+ \left[f'(x)  -f(x) \frac{g'(x)}{g(x)}\right] 
\right\}
\label{eq:Sx-1valpha}
\end{equation}
where we have 
separated the contribution of the total derivative.
This contribution gives the boundary factor $[g(x(t_0))/g(x(t_f))]^{1/2}$, that together with the measure $(\det(g^2))^{-1/2}$
fixes the normalization of the path integral.
If
$(\det(g^2))^{-1/2}$ is understood as $\prod_\ell[g(x_\ell)g(x_{\ell+1})]^{-1/2}$ on a time lattice,
the product of  the two factors  is the end-point normalization $\prod_\ell1/g(x_{\ell+1})$ used in
Ref.~\cite{Cugliandolo_2019}.
It is not difficult to confirm that Eq.~(\ref{eq:Sx-1valpha}) is covariant under the transformation given by Eqs.~(\ref{eq:U}),~(\ref{eq:F-main}) and (\ref{eq:G-main}).  We can also prove that the measure satisfies 
\begin{equation}
\frac{{\cal D}x }{g(x)}= \frac{{\cal D}u}{ G(u) }
\label{eq:measure}
\end{equation}
where the denominator in the left-hand-side is due to the factor $({\rm det} (g^2))^{-1/2}$ in Eq.~(\ref{eq:ZIxi}). 

Finally, making the identification $g\to \sqrt{2D} g$ we obtain, 
\begin{align}
S[x] =
&\int  {\rm d}t \;  \left\{ \frac{1}{4D} \left[ \frac{\dot x-f(x)}{g(x)}\right]^2 
+\frac{1}{2}\left[f'(x)  -f(x) \frac{g'(x)}{g(x)}\right] \right\}
\; , 
\label{eq:Sx-1vStr}
\end{align} 
which exactly coincides with the action presented in Ref.~\cite{Cugliandolo_2019}. The latter was obtained 
with a direct construction of the Onsager--Machlup action (no fermions)
using a higher-order discretization scheme, with a quadratic term in the infinitesimal increment of the stochastic process 
(see Eqs.~(\ref{eq:covariant-M-discretization}) and (\ref{eq:betag})). While quadratic terms are irrelevant at the level of the Langevin equation, they contribute non-trivially to the  generating functional in the vanishing time-step limit. The final result is precisely the action in Eq.~(\ref{eq:Sx-1vStr}) 
with the measure normalized as in Eq.~(\ref{eq:measure}).

In the continuum calculation, we fixed $G_R(0)=1/2$
and regularized the Grassmann contact terms by setting $\Theta_2=0$.
To relate this choice to the discretization of
Ref.~\cite{Cugliandolo_2019}, we interpret the action in
Eq.~(\ref{eq:Sx-1valpha}) in the $(1/2, \beta_g)$ discretization with $\beta_g$ given
by Eq.~(\ref{eq:betag}) and we expand it about the mid-point. The quadratic shift
$\beta_g(\Delta x)^2$ in the discretization point generates the local
potential
\[
\frac18(2g'^2-gg'')=\frac18 g'^2+\frac18(g'^2-gg'').
\]
The first term coincides with the contact contribution to the action
obtained from the Grassmann determinant on the mid-point lattice,
which  corresponds to the choice $\Theta_2=1/4$. The second
accounts for the change from mid-point to end-point normalization
of the measure. Thus, the continuum prescription $\Theta_2=0$
corresponds to reading the resulting Onsager--Machlup action
with the covariant $(1/2,\beta_g)$ discretization and the end-point
normalization of the measure.

To obtain the action in Eq.~(\ref{eq:Sx-1valpha}), we first integrated over the auxiliary time-dependent variables and then over the Grassmann variables {$\bar\xi,\xi$}. This order simplifies the Grassmann integration because the $\bar\xi\xi$ prefactors  do not contain the response field. Integrating in reverse order is possible but more complex, as some prefactors involve the response field, which is a non-regular function of time. We detail this alternative approach in the next subsection.

\subsection{Order of integration: Grassmann variables followed by the response variable}
\label{subsubsec:second-route}

In Sec.~\ref{subsubsec:first-route}, we  deduced the Onsager--Machlup action $S[x]$ in Eq.~(\ref{eq:Sx-1vStr}), which is written only in terms of the single variable $x$. 
For this, we first integrated over the response variable $\varphi$ and then over the Grassmann variables $\{\bar\xi,\xi\}$. 
Here,  we proceed in the opposite order. 
In this case, the integration on the Grassmann variables is more tricky  and it is important to clarify how to perform the 
calculation that leads to the same final result  for $S[x]$.

We can rewrite the  generating functional in the following compact way, 
\begin{equation}
Z=\int{\cal D}x {\cal D}\varphi \;  e^{-\mathop{\mathlarger{\int}} {\rm d}t \; \left[ {\rm i} \varphi\left(\dot x-f\right)+\frac{1}{2}g^2\varphi^2\right] }
\; I_{\xi}[\varphi,x]
\label{app:Z}
\end{equation}
where 
\begin{equation}
I_{\xi}[\varphi,x]=\int {\cal D}\bar\xi{\cal D}\xi \; e^{\mathop{\mathlarger{\int}} {\rm d}t \; \left( \bar\xi\dot \xi - {\rm i} \varphi g g'\bar\xi\xi-f'\bar\xi\xi \right) } \, .
\label{app:Ixi}
\end{equation}
Applying  the following change of variables 
\begin{equation}
\varphi=\tilde\varphi-\frac{{\rm i}}{g^2}\left(\dot x-f\right)
\end{equation}
 in Eq.~(\ref{app:Z}), the linear term in $\varphi$ is eliminated and we obtain
\begin{equation}
Z=\int{\cal D}x \; e^{-\frac{1}{2} \mathop{\mathlarger{\int}} {\rm d}t \frac{1}{g^2}\left(\dot x-f\right)^2}
\! \int{\cal D}\tilde\varphi  \; e^{-\frac{1}{2} \mathop{\mathlarger{\int}} {\rm d}t \; g^2\tilde\varphi^2}
I_{\xi}[\tilde\varphi,x]
\; , 
\label{app:Z2}
\end{equation}
where
\begin{equation}
I_{\xi}[\tilde\varphi,x]=\int {\cal D}\bar\xi{\cal D}\xi \; 
e^{\mathop{\mathlarger{\int}} {\rm d}t \; \left( \bar\xi\dot \xi - {\rm i} \tilde\varphi g g'\bar\xi\xi-k \bar\xi\xi \right) }
\; , 
\label{app:Ixi2}
\end{equation}
with $k(x(t))$ the same function defined in Eq.~(\ref{eq:kdef1}),
\begin{equation}
k(x(t))\equiv \frac{g'(x(t))}{g(x(t))}\left[\dot x(t)-f(x(t))\right]+f'(x(t)) \, .
\label{eq:k2}
\end{equation}

In order to explicitly perform the integration over the Grassmann variables, 
we Taylor expand the exponential, 
\begin{align}
&I_{\xi}[\tilde\varphi,x]=\sum_{n=0}^{\infty} \frac{(-1)^n}{n!}\int {\cal D}\bar\xi{\cal D}\xi \; e^{\mathop{\mathlarger{\int}} {\rm d}t \; \bar\xi(t)\dot \xi(t)} 
\left( \int dt \left[{\rm i} \tilde\varphi(t) g(t) g'(t)+k(t) \right]\bar\xi(t)\xi(t) \right)^n 
\!\! .
\end{align}
To see the structure of the integration, it is convenient to  write explicitly  the first few terms of the series 
\begin{align}
& 
I_{\xi}[\tilde\varphi,x]
=\int {\cal D}\bar\xi{\cal D}\xi \; e^{\mathop{\mathlarger{\int}} {\rm d}t \; \bar\xi(t)\dot \xi(t)} 
 \left\{1 -\int {\rm d}t_1 \; k_1\; \bar\xi_1\xi_1 
+\frac{1}{2!}\int {\rm d}t_1{\rm d}t_2 \;  k_1k_2 \;\bar\xi_1\xi_1 \bar\xi_2\xi_2 \right.
\nonumber \\ 
& 
 \ \ \
  -\frac{1}{3!}\int {\rm d}t_1{\rm d}t_2{\rm d}t_3 \;  k_1k_2k_3\; \bar\xi_1\xi_1 \bar\xi_2\xi_2\bar\xi_3\xi_3 
 +\frac{1}{4!}\int {\rm d}t_1{\rm d}t_2{\rm d}t_3{\rm d}t_4 \;  k_1k_2k_3k_4\; \bar\xi_1\xi_1 \bar\xi_2\xi_2\bar\xi_3\xi_3
 \bar\xi_4\xi_4 \nonumber \\ 
& 
\ \ \
-\frac{1}{2!}\int {\rm d}t_1{\rm d}t_2 \; \tilde\varphi_1\tilde\varphi_2 g_1 g'_1g_2 g'_2\; \bar\xi_1\xi_1\bar\xi_2\xi_2 
+\frac{3}{3!}\int {\rm d}t_1{\rm d}t_2{\rm d}t_3 \; k_1\tilde\varphi_2\tilde\varphi_3 g_2 g'_2g_3 g'_3\; \bar\xi_1\xi_1\bar\xi_2\xi_2
\bar\xi_3\xi_3 
\nonumber\\
&
\ \ \
-
\frac{6}{4!}  \int {\rm d}t_1{\rm d}t_2{\rm d}t_3{\rm d}t_4 \; k_1k_2\tilde\varphi_3\tilde\varphi_4 g_3 g'_3g_4 g'_4\; \bar\xi_1\xi_1\bar\xi_2\xi_2
\bar\xi_3\xi_3 \bar\xi_4\xi_4 
\nonumber\\
&
\ \ \
+\frac{1}{4!}  \int {\rm d}t_1{\rm d}t_2{\rm d}t_3{\rm d}t_4 \; \tilde\varphi_1\tilde\varphi_1 \tilde\varphi_3\tilde\varphi_4 
g_1 g'_1g_2 g'_2 g_3 g'_3g_4 g'_4\; \bar\xi_1\xi_1\bar\xi_2\xi_2
\bar\xi_3\xi_3 \bar\xi_4\xi_4 + \dots \Big\}
\label{eq:expansion}
\end{align}
where, to make the expression compact, 
we introduced the notation $k_i\equiv k(x(t_i))$, $\tilde\varphi_i\equiv \tilde\varphi(t_i)$,  $\xi_i\equiv \xi(t_i)$,  $\bar\xi_i\equiv \bar\xi(t_i)$ and $g_i=g(x(t_i))$,  with $i=1,\ldots,n$.  We have only 
kept  even terms in powers of $\tilde\varphi$ since the odd ones vanish under integration over the Gaussian $\tilde\varphi$ measure. In addition, we ordered the terms,  placing first the ones with only $k$ factors (no $\tilde\varphi$ variables), while the rest of the terms are ordered in even powers of $\tilde\varphi$. These two classes of terms  (with and without $\tilde\varphi$) behave differently as we describe bellow. 

Now, we use the fact that~\cite{zinn-justin_quantum_2002}
\begin{equation}
\int {\cal D}\bar\xi{\cal D}\xi \; e^{\mathop{\mathlarger{\int}} {\rm d}t \; \bar\xi(t) \dot \xi(t)} \; \bar\xi(t)\xi(t')= \det({\rm d}/{\rm d}t) G_R(t-t')
\end{equation}
where $G_R(t)=\Theta(t)$ is the retarded Green's function of the operator ${\rm d}/{\rm d}t$
(similarly to what we have used in the derivation in Sec.~\ref{subsubsec:first-route}).

To compute the  correlation functions, we use (fermionic) Wick's theorem. The first three terms read, 
\begin{align}
\langle\bar\xi(t)\xi(t)\rangle&=G_R(0) 
\; , 
\label{eq:1G} \\
 \langle\bar\xi(t_1)\xi(t_1)\bar\xi(t_2)\xi(t_2)\rangle&= \det
\begin{pmatrix}
\langle \bar\xi(t_1) \xi(t_1) \rangle & \langle \bar\xi(t_1) \xi(t_2) \rangle
\\
\langle \bar\xi(t_2) \xi(t_1) \rangle & \langle \bar\xi(t_2) \xi(t_2) \rangle
\end{pmatrix}
\nonumber\\
&=G^2_R(0)-G_R(t_2-t_1)G_R(t_1-t_2) 
\; , 
 \label{eq:2G}  \\
 \langle\bar\xi(t_1)\xi(t_1)\bar\xi(t_2)\xi(t_2)\bar\xi(t_3)\xi(t_3)\rangle&= \det
\begin{pmatrix}
\langle \bar\xi(t_1) \xi(t_1) \rangle & \langle \bar\xi(t_1) \xi(t_2) \rangle & \langle \bar\xi(t_1) \xi(t_3) \rangle
\\
\langle \bar\xi(t_2) \xi(t_1) \rangle & \langle \bar\xi(t_2) \xi(t_2) \rangle & \langle \bar\xi(t_2) \xi(t_3) \rangle
\\
\langle \bar\xi(t_3) \xi(t_1) \rangle & \langle \bar\xi(t_3) \xi(t_2) \rangle & \langle \bar\xi(t_3) \xi(t_3) \rangle
\end{pmatrix}
\nonumber\\
&=G^3_R(0)-G_R(0)\left\lbrace G_R(t_2-t_1)G_R(t_1-t_2)\right.
 \nonumber \\
& \left. + G_R(t_3-t_1)G_R(t_1-t_3)+G_R(t_3-t_2)G_R(t_2-t_3)\right\rbrace \nonumber \\ 
&+ G_R(t_1-t_2)G_R(t_2-t_3)G_R(t_3-t_1) \nonumber \\
&+ G_R(t_2-t_1)G_R(t_3-t_2)G_R(t_1-t_3) 
\; . 
\label{eq:3G}
 \end{align}

The case of the two-point correlation function,  Eq.~(\ref{eq:2G}),  has already been discussed in  Eq.~(\ref{eq:GR2}).
In the  three point correlation,  Eq.~(\ref{eq:3G}), we distinguish the case with two equal times  from the case with all times  being different.   
For any pair of equal times $t_i=t_j$,   it vanishes 
because of $\xi^2(t)=(\bar\xi(t))^2=0$.  
The case  with all three times different  reads
 \begin{equation}
  \langle\bar\xi(t_1)\xi(t_1)\bar\xi(t_2)\xi(t_2)\bar\xi(t_3)\xi(t_3)\rangle=G_R^3(0)
  \end{equation}
since $G_R(t)G_R(-t)=0$ for $t\neq 0$
and one cannot simultaneously have $t_1> t_2>t_3>t_1$ or $t_2>t_1>t_3>t_2$.

Summarizing, the correlation function with $n$ pairs of $\bar\xi\xi$ is composed of  $n!$ terms,  each of them being a product of $n$ retarded Green's function.  The structure of these terms is very simple: 
\begin{equation}
\langle\bar\xi(t_1)\xi(t_1)\bar\xi(t_2)\xi(t_2)\ldots\bar\xi(t_n)\xi(t_n)\rangle=
\left\{
\begin{array}{lcl}
        G^n_R(0) & \mbox{if} & t_i\neq t_j, \forall \, i\neq j \\
        & & \\
        0 & \mbox{if} &  t_i=t_j \mbox{~for any~}  i\neq j 
    \end{array}
\right.
\label{eq:FermionicCorrelations}
\end{equation}
for $i,j=1,\ldots,n$.
The first line in this equation is a consequence of the retarded character of the Green's function
and  the second line is due to the anticommuting  character of the Grassmann variables.
 Equation~(\ref{eq:FermionicCorrelations}) can be checked explicitly 
in Eqs.~(\ref{eq:1G}), (\ref{eq:2G}) and (\ref{eq:3G}) for $n=1,2,3$, and it is not difficult to verify that it is satisfied for all $n$.

Now, to explicitly perform  the integrations, we notice two types of terms: with and without $\tilde\varphi$ fields.  The first four terms of 
Eq.~(\ref{eq:expansion}) depend on $k$ factors only and read
\begin{align}
&
\langle 1 \rangle -\int {\rm d}t_1 \; k_1\; \langle\bar\xi_1\xi_1 \rangle
+\frac{1}{2!}\int {\rm d}t_1{\rm d}t_2 \;  k_1k_2 \;\langle\bar\xi_1\xi_1 \bar\xi_2\xi_2 \rangle \nonumber \\
& \quad 
-\frac{1}{3!}\int {\rm d}t_1{\rm d}t_2{\rm d}t_3 \;  k_1k_2k_3\; \langle\bar\xi_1\xi_1 \bar\xi_2\xi_2\bar\xi_3\xi_3\rangle 
 +\ldots
\end{align} 

As in Sec.~\ref{subsubsec:first-route}, the coincident-time contributions of these terms are
not negligible, because $k$ is not a regular function. We treat them with the same prescription,
$\Theta_2=0$. Then, the contribution of the first $k$-dependent terms becomes
\begin{align}
1 & -G_R(0)\int {\rm d}t \; k(t)
+ \frac{1}{2!} \left(G_R(0)\int {\rm d}t \; k(t)\right)^2-\frac{1}{3!} \left(G_R(0)\int {\rm d}t \; k(t)\right)^3
 +\ldots
\end{align} 
The same happens with all higher order terms containing only products of $k$ functions.  
In this way, we can re-exponentiate the infinite series obtaining   
\begin{align}
\label{eq:k-reexponenciation}
I_{\xi}[\tilde\varphi,x] = & \;\; \det({\rm d}/{\rm d}t)\; e^{-G_R(0)\int {\rm d}t \; k(t)} \\
& 
-\frac{1}{2!}\int {\rm d}t_1{\rm d}t_2 \;  g_1 g'_1g_2 g'_2\;\tilde\varphi_1\tilde\varphi_2\langle \bar\xi_1\xi_1\bar\xi_2\xi_2\rangle \nonumber\\
&
+\frac{3}{3!}\int {\rm d}t_1{\rm d}t_2{\rm d}t_3 \; k_1 g_2 g'_2g_3 g'_3\;\tilde\varphi_2\tilde\varphi_3\langle \bar\xi_1\xi_1\bar\xi_2\xi_2
\bar\xi_3\xi_3 \rangle
\nonumber\\
&
- 
\frac{6}{4!}  \int {\rm d}t_1{\rm d}t_2{\rm d}t_3{\rm d}t_4 \; k_1k_2 g_3 g'_3g_4 g'_4\; \tilde\varphi_3\tilde\varphi_4\langle \bar\xi_1\xi_1\bar\xi_2\xi_2
\bar\xi_3\xi_3 \bar\xi_4\xi_4 \rangle
\nonumber\\
&
+\frac{1}{4!}  \int {\rm d}t_1{\rm d}t_2{\rm d}t_3{\rm d}t_4 \;  
g_1 g'_1g_2 g'_2 g_3 g'_3g_4 g'_4\; \tilde\varphi_1\tilde\varphi_2 \tilde\varphi_3\tilde\varphi_4\langle\bar\xi_1\xi_1\bar\xi_2\xi_2
\bar\xi_3\xi_3 \bar\xi_4\xi_4 \rangle+  O\left( \tilde\varphi^6\right)
\nonumber
\end{align}
where the fermionic correlations are given by Eq.~(\ref{eq:FermionicCorrelations}).

To complete the integration,  we are left with all  terms containing an even power of response variable $\tilde\varphi$.   
The response variable is  not a regular function of $t$. As we are working with Gaussian white noise, the integration over $\tilde\varphi$ has a local Gaussian measure.  This means that $\langle \tilde\varphi(t)\tilde\varphi(t')\rangle$ is delta correlated in time (times a $g$ dependent factor):
\begin{equation}
\int {\cal D}\tilde\varphi \; e^{-\frac{1}{2}\int {\rm d}t \; g^2 \tilde\varphi^2}\tilde\varphi(t)\tilde\varphi(t')
\propto
 (\det(g^2))^{-1/2} \, \frac{1}{g^2(x(t)) }\delta(t-t') \; 
 \label{eq:varphiCorrelation}
\end{equation}
ignoring the numerical factor.   As a consequence,  we are only left with integrals in which times have been 
forced to be equal two by two.  Therefore,  the terms of the series containing $\tilde\varphi$ fields cannot be re-exponentiated.
Their treatment below requires no prescription: on a time lattice, the Grassmann integral at
fixed $\tilde\varphi$ is linear in each $\tilde\varphi(t_\ell)$, so these terms vanish exactly. The
prescription for the $k$-dependent terms must therefore be applied after the integration over
$\tilde\varphi$, as done here. Applying it to the Grassmann determinant at fixed $\tilde\varphi$ would
restore the coincident-time term $\frac18\int{\rm d}t\,g'^2$ through the fluctuations of
$\tilde\varphi$.

Let us investigate these terms more carefully.   We can use Eq.~(\ref{eq:varphiCorrelation}) to functionally  integrate over $\tilde\varphi$ in  the $n=2$ term (the second term of Eq.~(\ref{eq:k-reexponenciation})).  We have 
\begin{align}
&\int {\rm d}t_1{\rm d}t_2 \; g_1 g'_1g_2 g'_2\; \langle\tilde\varphi_1\tilde\varphi_2\rangle\langle \bar\xi_1\xi_1\bar\xi_2\xi_2\rangle =  \nonumber \\
&=\int {\rm d}t_1{\rm d}t_2  \; (\det(g^2))^{-1/2}  (g'(t_1))^2\delta(t_1-t_2) \left[G^2_R(0)-G_R(t_2-t_1)G_R(t_1-t_2)\right] \nonumber \\
&=0
\; .
\end{align}
We clearly observe that the terms with $t_1=t_2$ cannot be discarded {\it a priori}, since they are the only contribution to the two-time
integral due to the delta-correlated response variable. However, in the end they force the integral to vanish identically because of the 
fermionic correlations.

The third term in Eq.~(\ref{eq:k-reexponenciation}) behaves in the same way, 
\begin{align}
\int {\rm d}t_1{\rm d}t_2{\rm d}t_3 & \; k_1 g_2 g'_2g_3 g'_3\;\langle\tilde\varphi_2\tilde\varphi_3\rangle\langle \bar\xi_1\xi_1\bar\xi_2\xi_2
\bar\xi_3\xi_3 \rangle \nonumber \\
& = \int {\rm d}t_1{\rm d}t_2{\rm d}t_3 \; k_1 (\det(g^2))^{-1/2}  (g'(t_2))^2\delta(t_2-t_3)
\langle \bar\xi_1\xi_1\bar\xi_2\xi_2
\bar\xi_3\xi_3 \rangle
\nonumber\\
& = 0
\; . 
\end{align}

Higher order terms are  computed by using Wick's theorem for the $\tilde\varphi$ correlations.   For instance, 
\begin{align}
\langle\tilde\varphi_1\tilde\varphi_2 \tilde\varphi_3\tilde\varphi_4\rangle=\langle\tilde\varphi_1\tilde\varphi_2 \rangle\langle\tilde\varphi_3\tilde\varphi_4 \rangle+\langle\tilde\varphi_1\tilde\varphi_3 \rangle\langle\tilde\varphi_2\tilde\varphi_4 \rangle+
\langle\tilde\varphi_1\tilde\varphi_4 \rangle\langle\tilde\varphi_2\tilde\varphi_3 \rangle
\label{eq:varphi-Wick}
\end{align}
where each term on the right-hand side is a product of Dirac delta functions.
Consequently, 
\begin{equation}
\int {\rm d}t_1{\rm d}t_2{\rm d}t_3{\rm d}t_4 \;
g_1 g'_1g_2 g'_2 g_3 g'_3g_4 g'_4\;  \langle\tilde\varphi_1\tilde\varphi_2 \tilde\varphi_3\tilde\varphi_4 \rangle\langle\bar\xi_1\xi_1\bar\xi_2\xi_2
\bar\xi_3\xi_3 \bar\xi_4\xi_4 \rangle=0 \, ,
\end{equation}
where we have used Eqs.~(\ref{eq:FermionicCorrelations}), (\ref{eq:varphiCorrelation}) and~(\ref{eq:varphi-Wick}).
Accordingly,  all other  higher order terms with even powers of $\tilde\varphi$ cancel after integration over $\tilde\varphi$. 
In short, the reason is that  the integration over  $\tilde\varphi$ projects the time integral domain into  regions with  $t_i=t_j$,  for any $i\neq j$,  due to the white character of the noise.   In these domains  the Grassmann correlation function is identically zero due to the anticommutation properties of these variables (see Eq.~(\ref{eq:FermionicCorrelations})).  

Therefore, we obtain 
\begin{align}
\int{\cal D}\tilde\varphi  \; e^{-\frac{1}{2} \mathop{\mathlarger{\int}} {\rm d}t \; g^2\tilde\varphi^2}
I_{\xi}[\tilde\varphi,x]= \det({\rm d}/{\rm d}t)\; e^{-G_R(0)\int {\rm d}t \; k(t)}  \, .
\label{eq:varphi-Integration}
\end{align}
Now, by replacing Eq.~(\ref{eq:varphi-Integration}) into Eq.~(\ref{app:Z2}) and using the explicit definition of $k(t)$, Eq~(\ref{eq:k2}),  we finally obtain
\begin{equation}
Z=\int{\cal D}x \; ({\rm det}(g^2))^{-1/2} \; e^{-S[x]}
\label{app:Zx-1v}
\end{equation}
with 
\begin{align}
S[x]=&\int  {\rm d}t \; \left\{ \frac{1}{2}\left(\frac{\dot x-f(x)}{g(x)}\right)^2 +\frac{1}{2} \left[f'(x)  -f(x) \frac{g'(x)}{g(x)}\right] 
+ \frac{1}{2} \; \dot x \frac{g'(x)}{g(x)} \right\}
\label{app:Sx-1v}
\end{align}
where we have fixed $G_R(0)=\alpha=1/2$. Equation~(\ref{app:Sx-1v}) coincides with Eq.~(\ref{eq:Sx-1v}).

We have thus shown that the covariant Onsager--Machlup action emerges independent of the integration sequence.

\section{Summary and conclusions}

In this work, we have revisited the fermionic path-integral
formulation of overdamped Langevin processes with multiplicative
white noise, with the aim of constructing a manifestly covariant
generating functional under non-linear coordinate transformations.
By employing the Grassmann-variable representation of the functional
Jacobian determinant, we have developed a purely continuous-time
framework that clarifies how covariance emerges from the interplay
between the non-differentiability of Wiener trajectories and the
state-dependent diffusion.

We first provided a detailed construction of the generating
functional for a single variable or scalar stochastic process,
explicitly introducing auxiliary commuting (response) and
anticommuting (Grassmann) fields.
One advantage of introducing anticommuting variables is that it
allows one to implement symmetries, with important physical
consequences, in a linear way. In particular, in equilibrium
systems, supersymmetric transformations which mix commuting
and anticommuting variables, are closely related to fundamental
properties such as probability conservation, fluctuation-dissipation
relations, and the underlying time-reversal symmetry of the
stochastic dynamics.

The resulting action, Eq.~(\ref{eq:action}), makes the
multiplicative nature of the noise explicit through derivative
couplings between the diffusion coefficient and the auxiliary
fields. Crucially,
using
ordinary calculus, we have demonstrated the covariance of the
complete fermionic generating functional.
Under a generic invertible change of variables $u=U(x)$, the action retains its form and the functional
measure is invariant,
as prescribed by
Eqs.~(\ref{eq:covariant-action})-(\ref{eq:covariant-measure}).
This shows that the fermionic path integral respects the same
transformation properties as the underlying
Langevin dynamics in the Stratonovich prescription,
without invoking explicitly any specific time discretization at the level
of the action.

After explicitly integrating out the auxiliary fields, we
recovered the Onsager--Machlup action with its covariant functional
measure. Our result provides a minor correction to that reported
in Ref.~\cite{arenas2010}. The integration was performed in two
distinct orders, first integrating the response field and then
the Grassmann variables, and vice versa, with careful handling
of the subtleties introduced by the white-noise character of
the fields. Both routes yield identical results, confirming
the internal consistency of the formalism, provided the same prescription for coincident-time
Grassmann contributions, $\Theta_2=0$, is used in both. This prescription is a
regularization: it is not implied by the Grassmann algebra.

The final Onsager--Machlup action coincides with the expression
previously derived using a higher-order (second-order in the
increment) discretization scheme~\cite{Cugliandolo_2019}.
This agreement demonstrates that the quadratic terms required
in a direct discrete-time construction are encoded in the regularization of the
Grassmann sector in a continuous-time approach, thereby bridging
regularization methods with functional techniques.
More precisely, regularizing the Grassmann contact terms with $\Theta_2=0$ 
 is equivalent to reading
the action with the $(\frac12,\beta_g)$ discretization and an end-point normalization of the
measure, in which case it describes the process correctly. The evaluation of the Grassmann integral
on the mid-point lattice would instead give the Stratonovich-discretized action, which is also correct
but not covariant. The continuum fermionic formulation therefore contains an implicit
discretization, encoded in the prescriptions of the Grassmann sector.

The continuum fermionic representation developed here is obtained
through the MSRJD construction starting from the overdamped Langevin equation,
with the functional Jacobian represented by Grassmann fields.
A recent perturbative algebraic field theory study has independently established
the correspondence between stochastic differential equations and
the MSRJD formalism for additive-noise and driftless multiplicative-noise
processes \cite{BonicelliDappiaggiDrago2026}. This result provides
further support for the underlying MSRJD construction, although it
does not directly establish the equivalence of the specific
fermionic representation considered here with the Langevin dynamics.

The agreement with the independently derived Onsager--Machlup action
provides an additional  non-trivial consistency check, although it does not
constitute a general proof of equivalence between the continuum
fermionic functional integral and arbitrary time-sliced
representations. In particular, a direct verification that the
short-time moments of a discretized fermionic kernel reproduce
the corresponding Langevin or Fokker--Planck generator, following
the criterion of Ref.~\cite{Ding2022}, remains an interesting question for
future investigation.

While the Stratonovich prescription is particularly convenient
for path-integral formulations, as it permits the use of ordinary
calculus, the consistent extension of the continuum
fermionic calculations and covariance proof to other stochastic
prescriptions within the $\alpha$-family remains an open question
(see~\cite{Lecomte-unpublished}). We expect that the fermionic
framework developed here can be extended to such cases, and we
plan to address this generalization in future work. Similarly,
the generalization to $d$-dimensional stochastic processes
achieved in~\cite{Thibaut-2022} using a higher order discretization
remains to be done with fermionic methods.

In summary, we have presented a continuous-time fermionic path-integral formulation
for overdamped scalar Langevin equations with multiplicative
white noise, establishing the covariance of the complete
generating functional under general non-linear transformations
using the standard chain rule and the convention $G_R(0)=1/2$.
This approach provides a useful framework
for further studies of symmetries, fluctuation theorems, and
non-equilibrium field theories.

\section*{Acknowledgments}
ZGA and DGB would like to thank the  Laboratoire de Physique Théorique et Hautes Energies (LPTHE) at 
Sorbonne Université for the warm hospitality during their stay as visiting Professors. 
The Brazilian agencies Conselho Nacional de Desenvolvimento 
Cient\'{\i}fico e Tecnol\'{o}gico (CNPq), Funda{\c{c}}{\~{a}}o de Amparo {\`{a}} 
Pesquisa do Estado do Rio de Janeiro (FAPERJ) and Coordena\c c\~ao de 
Aperfei\c coamento de Pessoal de N\'\i vel Superior (CAPES) are acknowledged for
partial financial support.  
DGB also acknowledge the support of the INCT project Advanced
Quantum Materials, involving the Brazilian agencies CNPq (Proc.
408766/2024-7), FAPESP (Proc. 2025/27091-3),   and CAPES.
LFC acknowledges financial support from the ANR-20-CE30-0031 grant THEMA.
We also acknowledge the support of  CAPES-Cofecub project PE3207237P.
We thank T. Arnoulx de Pirey, V. Lecomte and F. van Wijland for previous collaboration on 
this subject and very helpful discussions.

\appendix
\addtocontents{toc}{\fixappendix}

\section{Determinants and  Grassmann variables}
\label{sec:Determinants}

In order to make the paper self-contained, in this Appendix we briefly review some 
properties of Grassmann variables~\cite{arenas2010}.  
For a detailed treatment of this subject in statistical mechanics as well as in 
quantum field theory we refer the reader to Ref.~\cite{zinn-justin_quantum_2002}. 

We define a Grassmann variable $\theta$ and its conjugates $\bar\theta$. They satisfy  
\begin{equation}
\{\theta,\theta\}=\{\bar\theta,\bar\theta\}=\{\theta,\bar\theta\}=0 
\; , 
\end{equation}  
where $\{a,b\}=ab+ba$ is the anti-commutator of $a$ and $b$. This definition implies the nilpotent property 
$(\theta)^2=\bar\theta^2=0$. Therefore, any Taylor expandable function of these variables should be a polynomial of degree  $2$ at most: 
\begin{equation}
F(\theta,\bar\theta)=a + b\;\theta + c\;\bar\theta + d\;\bar\theta \theta
\label{Fn1}
\end{equation}  
The coefficients $a, b, c, d$ are  complex numbers. 

Differentiation with respect to the Grassmann variable is also a nilpotent operator $\partial^2/\partial\theta^2=0$ satisfying the Clifford algebra
\begin{equation}
\left\{\frac{\partial~}{\partial\theta},\frac{\partial~}{\partial\theta}\right\}=0
\; ,\;\;\; \quad
\left\{\frac{\partial~}{\partial\theta},\theta \right\}=1
\; ,  
\end{equation}  
and their conjugates. 

The two kinds of Grassmann variables $\theta$ and $\bar \theta$ are independent in the sense that 
$\partial \theta/\partial \bar \theta = \partial \bar \theta/\partial \theta =0$
Instead, $\partial \theta/\partial \theta =1$
and $ \partial \bar \theta/\partial \bar \theta =1$.
 
Interestingly, integration in a Grassmann space is the same operation as differentiation. Therefore, the use of a derivative or an integral symbol is a matter of taste. For instance, taking into account Eq.~(\ref{Fn1}),  
\begin{equation}
\int d\theta d\bar\theta\; F(\theta, \bar\theta)=d
\label{integral}
\end{equation}
That means that the integration over $d\theta d\bar\theta$ picks up the coefficient of the term $\theta\bar\theta$, 
in the same way as does the derivative $\partial^2/\partial \theta \partial \bar \theta$.

Let us consider the Gaussian Grassmann integral 
\begin{equation}
I_G(A)=\int \prod_{ij} {\rm d}\theta_i {\rm d}\bar\theta_j \;\;
e^{\bar\theta_{i}  A_{ij} \theta_j }
\; , 
\label{IGA}
\end{equation}
where $A_{ij}$ are the elements of a square matrix ${\mathbb{A}}$ and the 
indices $i,j$ can represent the different discrete times, $t_i = i \Delta t$ with $i=1, \dots, K$,  at which the Grassmann variables are evaluated, $\theta(t_i) = \theta_i$ and 
$\bar \theta(t_i) = \bar\theta_i$. The Grassmann variables at different times are independent.
According to Eq.~(\ref{integral}), the integral is the coefficient of the term proportional to $\bar\theta_1\theta_1\bar\theta_2\theta_2\ldots\bar\theta_K\theta_K$. 
Therefore, expanding the exponential in Eq.~(\ref{IGA}) in a ``finite'' Taylor series, and reordering the terms taking into account the anti-commuting properties of the Grassmann variables, 
\begin{equation}
I_G(A)=\det \mathbb{A}
\; . 
\label{detA}
\end{equation}
This result should be compared with the output of a normal Gaussian integral over real variables that is $\sim {\det^{-1/2} \mathbb{A}}$. Therefore, 
in the same way that the inverse of an  $K\times K$ matrix determinant  can be represented as a Gaussian integral over a set of $K$ complex variables, the determinant itself can be represented as a Gaussian integral over a  set of $K$ Grassmann variables.

We can generalize now the case of a discrete set of Grassmann variables $\{\theta_1,\ldots\theta_K\}$, to an infinite set of continuous variables; {\it i.e.}, a Grassmann function $\xi(t)$.  In this case, we can generalize Eqs.~(\ref{IGA}) and~(\ref{detA}) to
\begin{equation}
\det A=\int \mathcal{D}\xi \mathcal{D}\bar{\xi} \ e^{\int {\rm d}t {\rm d}t'\;\bar{\xi}(t) A(t,t')\xi(t')}
\; , 
\end{equation} 
where $\det A$ is a functional determinant and $\mathcal{D}\xi \mathcal{D}\bar{\xi}$ are functional integrations over Grassmann variables.
$A(t,t')$ is the kernel of the functional $A$. 
This is the formula used in Eq.~(\ref{detgrassman}) to represent the functional Jacobian of $\delta\hat O(t)/\delta x(t')$, 
and in the multi-variable case the one  of $\delta\hat O(t)/\delta x(t')$.

\section*{References}


\providecommand{\newblock}{}

\end{document}